\documentclass[aps,prd,showpacs,preprintnumbers,amsfonts,amsmath,amssymb]{revtex4}

\usepackage{subfigure}
\usepackage{dcolumn} % Align table columns on the decimal point
\usepackage{graphicx}
\usepackage{bm} % Bold math

\begin{document}

% user-defined commands
\newcommand{\earth}{\oplus}
\newcommand{\sun}{\odot}
\newcommand{\be}{\begin{equation}}
\newcommand{\ee}{\end{equation}}
\newcommand{\bea}{\begin{eqnarray}}
\newcommand{\eea}{\end{eqnarray}}
\newcommand{\bean}{\begin{eqnarray*}}
\newcommand{\eean}{\end{eqnarray*}}
\newcommand{\tranom}{f}
\newcommand{\tranomA}{f_1}
\newcommand{\tranomB}{f_2}
\newcommand{\trlongA}{\Lambda_1}
\newcommand{\trlongB}{\Lambda_2}
\newcommand{\trlong}{\Lambda}
\newcommand{\semimaj}{a}
\newcommand{\semimajA}{a_1}
\newcommand{\semimajB}{a_2}
\newcommand{\pos}{\bm{r}}
\newcommand{\posA}{\bm{r}_1}
\newcommand{\posB}{\bm{r}_2}
\newcommand{\posmag}{r}
\newcommand{\posAmag}{r_1}
\newcommand{\posBmag}{r_2}
\newcommand{\ecc}{e}
\newcommand{\eccA}{e_1}
\newcommand{\eccB}{e_2}
\newcommand{\longperA}{\varpi_1}
\newcommand{\longperB}{\varpi_2}
\newcommand{\longper}{\varpi}
\newcommand{\node}{\Omega}
\newcommand{\incl}{i}

% for revtex4
\newcommand{\aj}{Astron. J.}
\newcommand{\apjl}{Astrophys. J. Lett.}
\newcommand{\aap}{Astron. and Astrophys.}
\newcommand{\araa}{Annu. Rev. of Astron. Astrophys.}
\newcommand{\mnras}{Mon. Not. R. Astron. Soc.}
\newcommand{\pasp}{Publ. Astron. Soc. Pac.}

\title{ Solar system constraints on the Dvali-Gabadadze-Porrati braneworld theory of gravity }

\author{James B. R. \surname{Battat} }
\email[]{jbattat@cfa.harvard.edu}
\author{Christopher W. \surname{Stubbs} }
\email[]{stubbs@physics.harvard.edu}
\altaffiliation{Department of Physics, Harvard University}
\affiliation{Department of Astronomy, Center for Astrophysics, Harvard University, Cambridge, MA 02138}

\author{John F. \surname{Chandler} }
\email[]{chandler@cfa.harvard.edu}
\affiliation{Center for Astrophysics, Harvard University, Cambridge, MA 02138}

\date{\today}

\begin{abstract}
A number of proposals have been put forward to account for the observed accelerating expansion of the Universe through modifications of gravity. One specific scenario, Dvali-Gabadadze-Porrati (DGP) gravity, gives rise to a potentially observable anomaly in the solar system: all planets would exhibit a common anomalous precession, $d\omega/dt$, in excess of the prediction of General Relativity.  We have used the \textsc{Planetary Ephemeris Program} (PEP) along with planetary radar and radio tracking data to set a constraint of $\left|d\omega/dt\right|_{obs} < 0.02$ arcseconds per century on the presence of any such common precession.  This sensitivity falls short of that needed to detect the estimated universal precession of $\left|d\omega/dt\right|_{DGP} = 5\times 10^{-4}$ arcseconds per century expected in the DGP scenario.  We discuss the fact that ranging data between objects that orbit in a common plane cannot constrain the DGP scenario. It is only through the relative inclinations of the planetary orbital planes that solar system ranging data have sensitivity to the DGP-like effect of universal precession.  In addition, we illustrate the importance of performing a numerical evaluation of the sensitivity of the data set and model to any perturbative precession.
\end{abstract}

%96.30.-t Solar system objects 
%04.50.Kd Modified theories of gravity 
%04.50.-h Higher-dimensional gravity and other theories of gravity
%04.80.-y Experimental studies of gravity
%04.00.00 General Relativity and gravitation
%45.50.Pk Celestial mechanics
%95.30.Sf Relativity + Gravitation

\maketitle

\section{Introduction}
Measurements of the apparent brightnesses of Type Ia supernovae provide strong evidence that the Universe is accelerating in its expansion \cite{riessAccelUniv1998,perlmutterAccelUniv1999}.  Although General Relativity (GR) with a non-zero cosmological constant can accommodate these observations, it is a major challenge for theoretical physics to explain the observed value of the energy density of the vacuum \cite{weinbergCCProblem1989,polchinskiCCProb2006}.  This cosmological constant problem has motivated the theoretical physics community to explore alternative descriptions of gravity that can account for the accelerating Universe without the need for dark energy.

One class of such theories includes the five-dimensional braneworld gravity developed by Dvali, Gabadadze and Porrati (DGP gravity) \cite{dgpTheory}, in which standard model interactions are constrained to a four-dimensional brane, embedded in a higher-dimensional bulk.  The graviton is also constrained to the brane on small scales, but over cosmological distances is free to explore the full higher-dimensional space, thereby weakening gravity on Gigaparsec length scales.  The appealing feature of this theory is its ability to produce cosmic acceleration without invoking dark energy.  The one free parameter in DGP gravity is the crossover scale $r_c$, above which gravity becomes five-dimensional.  In order to produce the observed acceleration rate of the Universe, $r_c \approx 5$ Gpc.   

The self-accelerating phase of DGP gravity has been shown to contain a ghost at the linearized level \cite{koyamaGhosts2007}.  This suggests that the self-accelerated phase may be unstable and therefore not self-consistent.  For the purposes of this work, we point out that the interpretation of the ghost mode is a subject of debate in the current literature \cite{deffayet2006,deffayet2007}, and that there is at least one way to preserve self-acceleration while avoiding the ghost instability \cite{deRhamCascadingDGP2007}.  We also note that the ``normal'' (non-accelerating) phase of DGP gravity, which does not possess a ghost \cite{koyamaGhosts2007}, also predicts a periapse precession of the same magnitude but opposite sign from the self-accelerating phase \cite{lueStarkman2003}, and is thus constrained by the current work.

Lue and Starkman \cite{lueStarkman2003} derived the remarkable result that on the scale of the solar system, DGP gravity predicts a universal rate of periapse precession for bodies in nearly circular orbits, independent of not only the orbital scale but also the mass $M$ of the central body:
\be
\label{dgprate}
\left|\frac{d\omega}{dt}\right| = \frac{3c}{8r_c} = 5\times 10^{-4} \left( \frac{r_c}{5\text{ Gpc}}\right)^{-1} \text{ arcseconds per century}
\ee
where $c$ is the speed of light.  Thus, upper limits on the rates of precession of the planetary perihelia set a lower limit on $r_c$, the scale above which the DGP graviton is free to explore the bulk.  To see that the precession rate is independent of the orbital semi-major axis and of $M$, consider the weak-field potential of any gravitational theory $V(r) = V_{N}(r) + \delta V(r)$, where $V_{N}(r)=-GM/r$ is the Newtonian potential, and $\delta V(r)$ is a perturbation such that $\epsilon \equiv \delta V(r)/V_{N}(r) << 1$.  For nearly circular orbits, the precession per orbit $\delta\omega$ can be computed from $\delta\omega = \pi r\left(r^2\left(r^{-1}\epsilon\right)^\prime\right)^\prime$ \citep{dgpAndMoon}, where $^\prime\equiv d/dr$.  The precession rate $d\omega/dt$ is found by dividing $\delta\omega$ by the orbital period $P$.  For a Keplerian orbit, $P = 2\pi r^3/\sqrt{GM}$.  In DGP gravity, if $r<<r_*$ then $\delta V(r) = c\sqrt{GMr}/r_c$ \citep{lueStarkman2003}, which leads to a precession rate that is independent of both $r$ and $M$.  We have made use of a new scale, $r_*\equiv\left(r_g r_c^2\right)^{1/3}$, below which the metric is dominated by the central body rather than by the effects of the cosmological flow \citep{lueStarkman2003}.  Here, $r_g \equiv 2GM/c^2$ is the gravitational radius of a mass $M$.  For the Sun, $M=2\times 10^{30}$~kg, $r_g\approx 3$~km, and so $r_* \approx 130$ pc.  Therefore all solar system bodies easily satisfy $r<<r_*$.  Iorio \cite{iorioOrbitalEffectsOfDGP2005} extended Equation \ref{dgprate} to second order in the orbital eccentricity $e$.  He showed that $e$ does not appear at first order and so for solar system bodies, the eccentricity contributes less than 1\% to $d\omega/dt$.  We ignore the contribution of the eccentricity to the precession in this work.  

Using solar system ephemeris constraints from \cite{pitjevaRelativisticEffects2005}, Iorio has undertaken a program \cite{iorioOrbitalEffectsOfDGP2005,iorioDGP2005,iorioDGP2006,iorioPrecessionConstraints2007} to constrain anomalous gravitational effects using solar system observations.  Our approach here is similar, but with three differences: (1) We use an independent ephemeris code, the \textsc{Planetary Ephemeris Program} (PEP) \cite{PEP_description} to arrive at limits on anomalous precession. (2) We undertake a constrained global fit to the motions in the solar system, as advocated by Nordvedt \cite{nordtvedtGrandSolarSystemFits}. (3) We perform a simulation to quantitatively assess our sensitivity to a DGP-like precession.  To test the DGP scenario, we constrain the anomalous precession rates of the planets (beyond the GR prediction) to be equal, as specified by Equation \ref{dgprate}.  

The recent renaissance of Lunar Laser Ranging (LLR), brought about by the millimeter-precision range measurements from the Apache Point Observatory Lunar Laser-ranging Operation (APOLLO, see \cite{APOLLOINSTR}), promises improved sensitivity to lunar orbital parameters.  As discussed in \cite{dgpAndMoon}, LLR observations could provide substantially improved tests of the DGP scenario.  As seen in Equation \ref{dgprate}, DGP theory predicts the same amount of periapse motion for the Moon as for the planets, even though they orbit different central bodies.  We will defer a consideration of improved lunar ranging limits on DGP to a future publication.  Our focus here is to explore DGP limits based on ranging between planets in orbit about the Sun.

\section{The PEP Ephemeris Model:\protect\\ Using Precessions to Link Observations with Theory}
PEP integrates the Parametrized Post-Newtonian (PPN) equations of motion for massive bodies \cite{willTEGP,WillNordtvedt1972}, with free parameters that describe the masses and orbital properties of solar system objects.  The equations of motion also encode deviations from standard physics that include, but are not limited to, the violation of the equivalence principle and the time rate of change of the gravitational constant.  The model parameter estimates are refined by minimizing the residual differences, in a weighted least-squares sense, between observations and model predictions.  This procedure can be iterated until convergence criteria are met.  

The PEP software has enabled constraints on departures from standard physics.  For example, it has been used to place limits on the PPN parameters $\beta$ and $\gamma$ (see e.g. \cite{reasenbergViking1979,CHANDLER1996}) and on possible violations of Lorentz Invariance \cite{battatSME2007}.  For the results described here we fixed the PPN parameters to their GR values ($\beta=\gamma=1$), and we modified the code to allow for a common perihelion precession for all planets in excess of the GR predictions.  This approach allows us to quantitatively assess the covariance between the anomalous precessions and the more mundane solar system parameters such as planet masses and initial orbital elements.

Our modification to the PEP code allows for independent limits on any anomalous precession of the planets in the solar system, or more generally, for constrained ratios of precession rates for the different solar system objects.  Rather than generating a specific version of PEP to encode the equations of motion for each speculative gravity scenario, we favor taking this perturbative approach based on anomalous precessions.  For the results described here we constrained all precession rates to be equal, as predicted by DGP gravity.  

For a broad class of extensions to GR, the PPN formalism \cite{WillNordtvedt1972,willTEGP} has been a very useful framework.  Unfortunately many of the interesting ideas that have been put forward recently (motivated by the Dark Energy crisis) do not readily map onto the PPN parametrization.  We are not aware of an analog of the PPN formulation that readily encompasses the diverse modified-gravity scenarios that are currently under consideration.  For example, even a simple Yukawa modification to GR does not naturally lend itself to a PPN description.  The ``forward modelling'' approach, to look for the (necessarily small) corrections to GR dynamics in the solar system in the form of perihelion precessions, provides a convenient tie point between observation and a broad class of speculative theories.

\begin{figure}
\includegraphics[width=\columnwidth,keepaspectratio,clip]{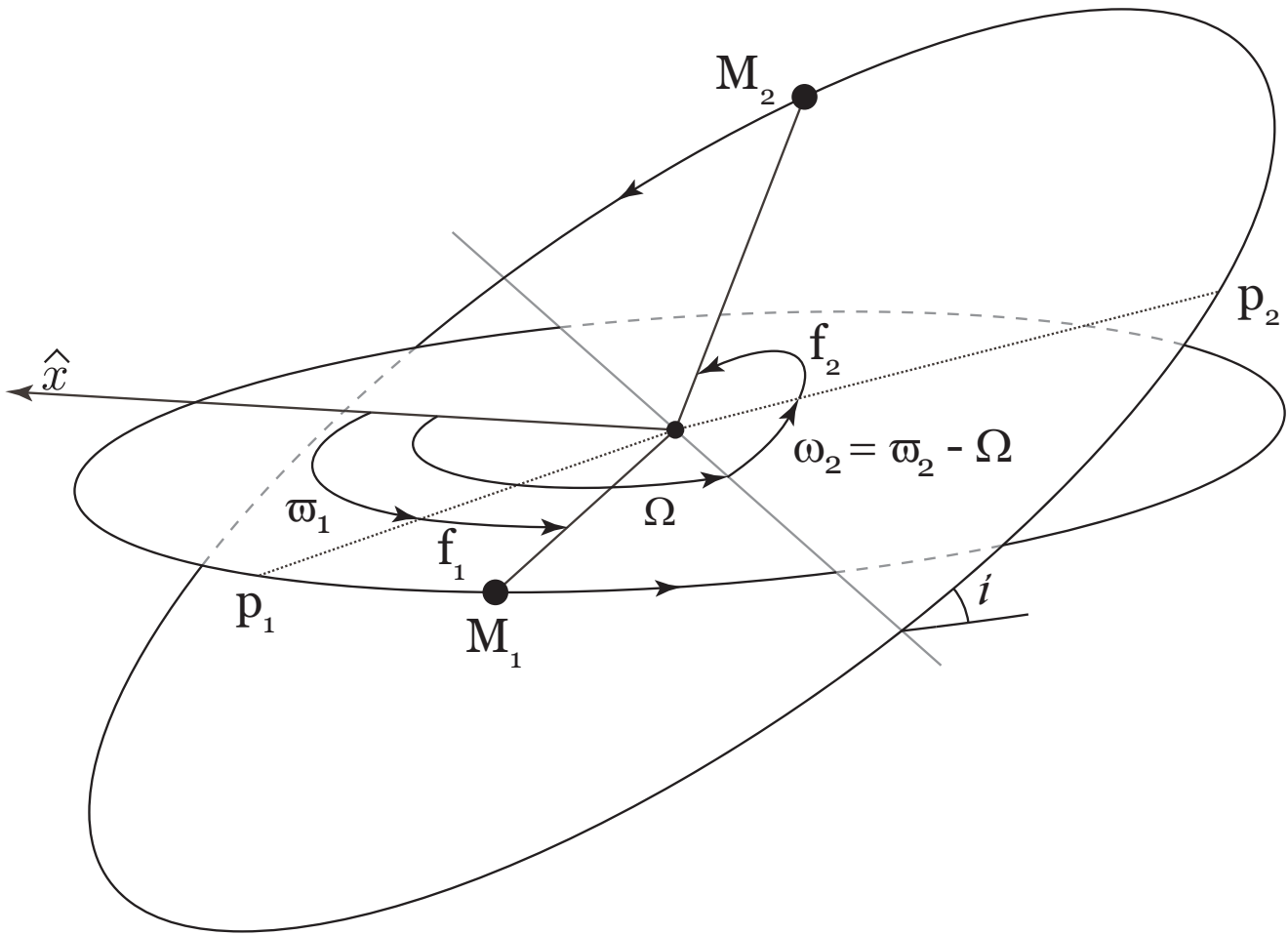}
\caption{\label{fig:orbitalGeometry}  Two planets, $M_1$ and $M_2$ (assumed massless), in Keplerian orbits about a common focus (e.g., the Sun).  The perihelia are labeled $p_1$ and $p_2$.  The longitudes of perihelion, $\longperA$ and $\longperB$ are measured with respect to the reference direction $\bm{\hat{x}}$, with $\longperB$ the algebraic sum of non-coplanar angles.  The true anomalies, $\tranomA$ and $\tranomB$, are measured with respect to their respective perihelia, and the true longitudes are defined by $\trlongA= \longperA+\tranomA$ and $\trlongB=\longperB+\tranomB$.  The inclined orbit has two additional parameters, the longitude of the ascending node, $\node$, and the inclination, $\incl$.  In our solar system, the planetary orbits are inclined by only a few degrees with respect to the ecliptic (the plane of the Earth's orbit about the Sun).}
\end{figure}

In fact, a bound, two-body system will exhibit closed orbits for only two central potentials:  $V(r)\propto r^2$ (e.g. Hooke's law), and $V(r)\propto 1/r$ (e.g. Newtonian gravity) \cite{landl}.  A perturbation to the Newtonian $1/r$ potential will, in general, violate Kepler's First Law by generating unclosed orbits.  These orbits are characterized by the rate of precession of the argument of periapse (or perihelion when the body orbits the Sun).  The argument of periapse $\omega$, defined in the orbital plane, is the angle between the ascending node and periapse (closest approach).  The argument of periapse should not be confused with the longitude of periapse, $\longper$, which is the sum of $\omega$ and the longitude of the ascending node $\node$.  In general, the angles $\omega$ and $\node$ are non-coplanar and so $\longper$ is often called the broken angle.  In Figure \ref{fig:orbitalGeometry}, $\omega_2 = \longperB-\node$, and $\omega_1=\longperA$ (the longitude of the ascending node for the orbit of mass 1 is undefined).  In a purely Keplerian orbit, $\omega$, $\node$ and $\longper$ do not change over time. 

Perihelion precession has been used before to distinguish theories of gravity.  For example, GR garnered early success from its ability to explain the long-standing 43 arcsecond per century discrepancy between the observed advance of Mercury's perihelion and the Newtonian prediction.  This difference is caused by the $1/r^3$ perturbation to the effective gravitational potential in the weak-field, slow-motion limit of GR \cite{hartleTextbook}.  Similarly, any modification to gravity or new fundamental interaction that deviates from a $1/r$ potential could be tested through its effect on the periapse of solar system bodies.  A useful framework for computing the expected precessions for a broad class of non-$1/r$ central potentials has been developed by Adkins and McDonnell \cite{adkinsMcdonnellPrecessions}.  We invite proponents of modified-gravity models to provide the expected ratios of periapse precessions in the solar system, and we would be pleased to undertake a PEP analysis at these fixed precession ratios. 

\section{DGP Precession, Global Rotation and the need for Inclined Orbits}
One might legitimately worry that a set of \textit{differential} observations taken within the solar system, namely the ranges between the Earth and other solar system bodies, might not be sensitive to a universal precession of the kind predicted by DGP gravity.  As we discuss below, differences in the inclinations of the planetary orbits with respect to the ecliptic are vital for testing DGP in the limit of small eccentricity, as they break the degeneracy between a DGP-like precession and a global rotation.

As described by Williams and Standish \cite{williamsStandishDynamicalReferenceFrames}, range data between two objects in heliocentric orbits can be used to determine the two orbital semi-major axes and the difference between the orbital angular frequencies of the bodies.  If the orbits are coplanar, then the ranging data are not sensitive to a common (DGP-like) precession of the perihelia of the two orbits, because the constant excess rotation rate vanishes from the orbital angular frequency difference.  In other words, a rotation preserves both the lengths and the relative angle between vectors.  Adding additional coplanar orbits and associated range data does not help.  Absent a determination of the phase of one of the objects in its orbit around the Sun, a DGP-like common precession would remain undetectable.  For non-coplanar orbits, however, range data between planets \textit{can} distinguish a global rotation from a DGP-like precession, given adequate measurement accuracy.  

To see that non-coplanar orbits are required to distinguish DGP gravity from a global rotation, consider the range, $\rho(t)$, between the centers of mass of two planets in orbit about the Sun with position vectors $\posA(t)$ and $\posB(t)$
\be
\label{dgprange}
\rho^2(t) = \posAmag^2 + \posBmag^{2} - 2\posA\cdot\posB
\ee
where $\posmag = |\pos|$ and the labels $1$ and $2$ refer to the reference body (e.g., the Earth) and the body on the inclined orbit (the ranged-to body), respectively.  In an elliptical Keplerian orbit, $r$ is related to the angle from periapse (called the true anomaly), $\tranom$, by 
\[
\posmag = \frac{\semimaj(1-\ecc^2)}{1+\ecc\cos\tranom}
\]
where $e$ is the eccentricity of the orbit, and $a$ is the orbital semi-major axis.  From the law of cosines in spherical trigonometry, the angle between the position vectors of the two bodies is
\begin{equation*}
\frac{\posA\cdot\posB}{\posAmag\posBmag} = \cos(\trlongB-\node)\cos(\trlongA-\node) + \sin(\trlongB-\node)\sin(\trlongA-\node)\cos \incl.
\end{equation*}
where $\trlongA=\longperA+\tranomA$ and $\trlongB=\longperB+\tranomB$ are the true longitudes of the bodies, $\longperA$ and $\longperB$ are the longitudes of periapse for the two bodies, $\node$ is the longitude of the ascending node and $i$ is the mutual orbital inclination.  Figure \ref{fig:orbitalGeometry} depicts the relevant geometry.  In an unperturbed Keplerian orbit, $\longper$, $\Omega$ and $i$ are static, and only $\tranom$, the angular position of the body, measured with respect to periapse, changes with time.  Periapse precession causes $\longper$ to change in time as well.

Following Williams and Standish \cite{williamsStandishDynamicalReferenceFrames}, we expand Equation \ref{dgprange} to first order in the orbital eccentricities and to second order in the inclination (the inclination appears only in even powers)
\be
\label{rangeexpansion}
\rho^2(t) = \semimajA^2 + \semimajB^2 - 2 \semimajA \semimajB \cos(\trlongB-\trlongA) + E + I
\ee
where
\bean
E \equiv &-& 2 \semimajA^2 \eccA \cos\tranomA 
             - 2 \semimajB^2 \eccB \cos\tranomB  \\*
    &+& 2 \semimajA\semimajB\left[\eccA\cos\tranomA\cos(\trlongB-\trlongA) \right. \\*
    &+&   \left.\eccB\cos\tranomB\cos(\trlongB-\trlongA)\right]
\eean
and
\[
I \equiv \frac{\semimajA\semimajB}{2} \sin^2\incl
            \left[\cos(\trlongB-\trlongA) - 
                  \cos(\trlongA+\trlongB-2\node)\right].
\]
To first order in the eccentricity and second order in the inclination, Equation \ref{rangeexpansion} is equivalent to Equation 3 of \cite{williamsStandishDynamicalReferenceFrames}, which is written in terms of mean anomalies and longitudes.  The first three terms in Equation \ref{rangeexpansion} simply give the range between circular, coplanar orbits.  The function $E$ accounts for the effect of the eccentricities of the orbits, and $I$ arises from the relative inclination of the orbits.  
It is immediately clear that the longitudes appear only through the difference, $\trlongB-\trlongA$, except in $I$.  Therefore measurements of $\rho(t)$ cannot distinguish between a DGP-like universal precession and a net rotation unless $i\ne0$.  Furthermore, the $\sin^2\incl$ term strongly suppresses the effect of a DGP-like precession on the observable range to planets whose orbits are only mildly inclined to the orbital plane of the reference planet (e.g., the ecliptic).

\section{DGP Perturbation to Planetary Ranges}
\begin{figure*}
\subfigure[Mercury]{\includegraphics[scale=0.4,keepaspectratio,clip]{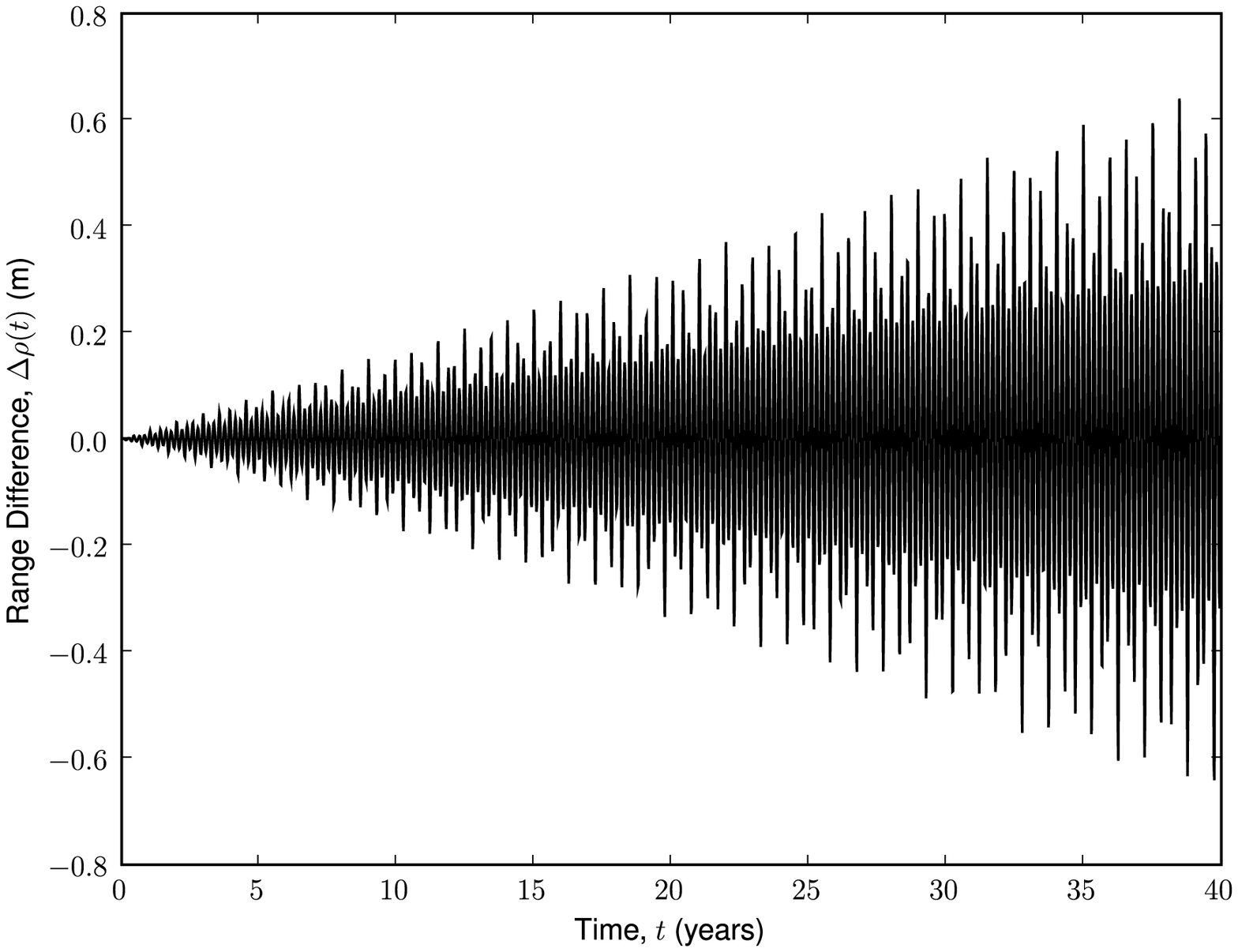}}
\subfigure[Mars]{\includegraphics[scale=0.4,keepaspectratio,clip]{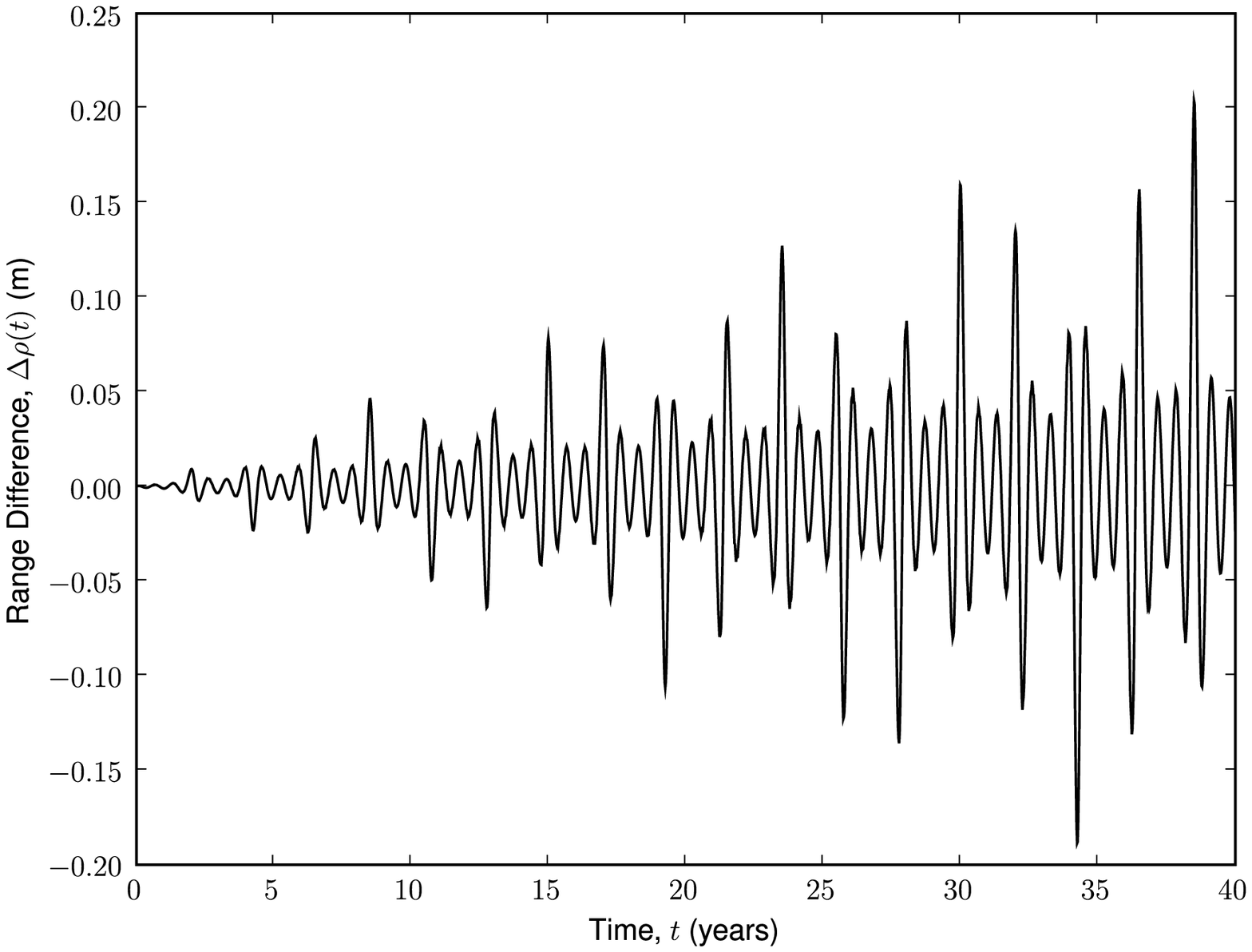}}
\caption{\label{fig:deltaRange} The estimated evolution of the range difference, $\Delta\rho(t)$, between DGP and Newtonian gravity to the planets Mercury (\textit{left}) and Mars (\textit{right}).  The difference of zero at the start of the time span reflects the freedom of the model to choose the initial anomaly $f$ of each planet independently.}
\end{figure*}

Given that the planetary orbits (save for the Earth's) in our solar system are inclined to the ecliptic, we can estimate the size of the DGP perturbation to interplanetary ranges.  Over a span of time, $\Delta t$, the total DGP-induced perihelion precession is $\Delta\omega \equiv (d\omega/dt)\Delta t << 1$.  To first order in $\Delta\omega$, the difference, $\Delta\rho(t)$, between the DGP range prediction and the Newtonian range prediction, $\rho_N(t)$ given in Equation \ref{rangeexpansion}, is
\be
\label{deltaRange}
\Delta\rho(t) = \frac{\semimajA \semimajB}{2\rho_N(t)}
                \Delta\omega
                \sin^2 \incl 
                \sin(\trlongA+\trlongB-2\node).
\ee
In our solar system, the planetary orbital inclinations with respect to the ecliptic are small, and therefore the $\sin^2\incl$ factor strongly suppresses the DGP signature.  For example, the orbits of Mercury and Mars are inclined to the ecliptic by 7.0 and 1.8 degrees, and so $\sin^2\incl = 10^{-2}$ and $10^{-3}$, respectively.  Unfortunately, there are no range measurements of Pluto, whose orbital inclination is 17 degrees (and so $\sin^2i \approx 0.1$).  Using Equation \ref{deltaRange}, we have computed $\Delta\rho(t)$ over 40 years for Mercury and Mars using the DGP prediction of $\left|d\omega/dt\right| = 5\times10^{-4}$ arcseconds per century.  The range differences, shown in Figure \ref{fig:deltaRange}, do not exceed one meter.  As described in Section \ref{observations}, the single-point measurement uncertainties in the ranging data to Mercury and Mars are approximately 200~m and 5--40~m, respectively.  

%%%%%%%%%%%%%%%%%%%%%%%%%% fig 1 %%%%%%%%%%%%%%%%%%%%%%%%%%%%%%%%%%
%%%%%%%%%%%%%%%%%%%%%%%%%%%%%%%%%%%%%%%%%%%%%%%%%%%%%%%%%%%%%%%%%%%%

%%%%%%%%%%%%%%%%%%%%%%%%%% fig 2 %%%%%%%%%%%%%%%%%%%%%%%%%%%%%%%%%%
%%%%%%%%%%%%%%%%%%%%%%%%%%%%%%%%%%%%%%%%%%%%%%%%%%%%%%%%%%%%%%%%%%%%

%J2        = 2.12  +/- 0.06
%domega/dt = 0.001 +/- 0.002 ``/cy

\section{Other sources of precession}
The planetary perihelia precess from Newtonian perturbations alone, by amounts much larger than relativistic effects.  For example, Newtonian perturbations from other planets induce a perihelion precession in Mercury's orbit of more than 500 arcseconds per century.  In addition, the Newtonian potential of the oblate Sun deviates from that of a spherically symmetric body, causing the perihelia of the planets to precess.  A planetary orbit that is inclined by an angle $\theta$ to the rotation axis of the Sun will experience the following rate of perihelion precession due to the Sun's mass quadrupole moment \citep{hartleTextbook}:
\be
\label{precJ2}
\left(\frac{d\omega}{dt}\right)_{J_{2{\sun}}} = \frac{6\pi}{P}\left(\frac{R_{\sun}}{a(1-e^2)}\right)^2 J_{2{\sun}}\left(\frac{1-3\cos^2\theta}{2}\right) \propto a^{-7/2}.%\frac{1}{P}.
\ee
Here $a$ is the orbital semi-major axis, $e$ is the orbital eccentricity, $P$ is the orbital period of the planet, $R_{\sun}$ is the mean radius of the Sun, and $J_{2{\sun}}$ is the dimensionless coefficient of the quadrupole moment of the Sun's mass distribution.

For comparison, the classic GR prediction for the perihelion precession of an orbiting planet is
\be
\label{precGR}
\left(\frac{d\omega}{dt}\right)_{GR} = \frac{6\pi G}{c^2}\frac{M_{\sun}}{a(1-e^2)}\frac{1}{P} \propto a^{-5/2}
\ee
where $M_{\sun}$ is the mass of the Sun.  This term, which has a different radial dependence than does the effect of the solar oblateness, produces the predicted 43 arcseconds per century perihelion motion for Mercury mentioned earlier.  The amounts of perihelion precession for each planet contributed by Equations \ref{precJ2} and \ref{precGR} are listed in Table \ref{tab:precSources}.

%%%%%%%%%%%%%%%%%%%%%%%%%%%%%%% Table %%%%%%%%%%%%%%%%%%%%%%%%%%%%%%%%%%%%%%%%
 \begin{table}%[H] add [H] placement to break table across pages
 \caption{\label{tab:precSources}The perihelion precession rates, in arcseconds per century, from GR and the solar oblateness, computed from Equations \ref{precJ2} and \ref{precGR} with $M_{\sun}=2\times 10^{30}$ kg, $R_{\sun}=6.95\times10^5$ km, $J_{2{\sun}} = 2\times10^{-7}$, and solar equatorial orbits ($\theta=\pi/2$) assumed for all bodies.  The DGP prediction for $r_c=5$ Gpc is 5$\times 10^{-4}$ arcseconds per century, which is less than the GR effect for $r<37$ AU, but exceeds the precession from the oblate Sun for planets outside the Earth's orbit.  As in the text, $a$ is the orbital semi-major axis (AU), and $P$ is the orbital period (years).}
 % Mass of Sun 2x10^30 kg
 % J2_sun = 2x10^-7
 % Rsun   = 695,000 km
 % Planet   a (AU)  e       P (yr) 
 % Mercury  0.387  0.2056   0.2408
 % Venus    0.723  0.0068   0.6152
 % Earth    1.000  0.0167   1.0000
 % Mars     1.524  0.0934   1.8809
 % Jupiter  5.2028 0.0483  11.8622
 % Saturn   9.5388 0.0560  29.4577
 % Uranus  19.1914 0.0461  84.0139
 % Neptune 30.0611 0.0097 164.793
 % Pluto   39.5294 0.2482 248.54
 \begin{ruledtabular}
 \begin{tabular}{lrrrr}
 Planet  &  $a$ & $P$ & $\left(\frac{d\omega}{dt}\right)_{J_{2{\sun}}}$ & $\left(\frac{d\omega}{dt}\right)_{GR}$ \\
 \hline
 Mercury &   0.39 &  0.24 & $2.5 \times 10^{-2}$ & 43.0                 \\
 Venus   &   0.72 &  0.62 & $2.6 \times 10^{-3}$ &  8.6                 \\
 Earth   &   1.00 &  1.00 & $8.4 \times 10^{-4}$ &  3.8                 \\
 Mars    &   1.52 &  1.88 & $1.9 \times 10^{-4}$ &  1.4                 \\
 Jupiter &   5.20 &  11.9 & $2.6 \times 10^{-6}$ & $6.2 \times 10^{-2}$ \\
 Saturn  &   9.54 &  29.5 & $3.1 \times 10^{-7}$ & $1.4 \times 10^{-2}$ \\
 Uranus  &   19.2 &  84.0 & $2.7 \times 10^{-8}$ & $2.4 \times 10^{-3}$ \\
 Neptune &   30.1 & 164.8 & $5.6 \times 10^{-9}$ & $7.8 \times 10^{-4}$ \\
 Pluto   &   39.5 & 248.5 & $2.4 \times 10^{-9}$ & $4.2 \times 10^{-4}$ \\
 \end{tabular}
 \end{ruledtabular}
 \end{table}
%%%%%%%%%%%%%%%%%%%%%%%%%%%%%%%%%%%%%%%%%%%%%%%%%%%%%%%%%%%%%%%%%%%%%%%%%%%%%

\section{Observations}
\label{observations}
There is a rich, publicly available archive of range and Doppler measurements to the terrestrial planets, spacecraft and landers.  Table \ref{tab:dataset} summarizes the data that were used in this analysis, including the type and duration of the observations and the typical associated measurement uncertainties.  

The range and Doppler measurements to Mercury and Venus were obtained by reflecting radar signals from the planets, and therefore depend on the planetary surface topography.  Similar radar measurements of Mars exist but have been supplanted by superior radio tracking to Mars orbiters and landers, most notably the Mars Global Surveyor (MGS) and Mars Odyssey orbiters.  Range observations to the Jovian planets are extremely limited.  There are four radio tracking observations of spacecraft as they passed by Jupiter (Pioneer 10 and 11 and Voyager 1 and 2) and two such observations for Saturn (Voyager 1 and 2).  No observations of Uranus, Neptune or Pluto were used here.  More detailed information about the planetary range and Doppler data sets is provided by Standish \cite{standishSSData1990}.  LLR data were not used in this work.  This allows for a more direct comparison with the previous results of Pitjeva \cite{pitjevaRelativisticEffects2005}, and the subsequent DGP interpretations by Iorio \cite{iorioDGP2005,iorioDGP2006,iorioPrecessionConstraints2007}.

%%%%%%%%%%%%%%%%%%%%%%%%%%%%% TABLE %%%%%%%%%%%%%%%%%%%%%%%%%%%%%%%%%%%%%%%%%%%%
\begin{table*}
\caption{\label{tab:dataset}Characteristics of the data used in this analysis.  $N_{obs}$ and $\sigma_{obs}$ are the number of observations and typical single-point measurement uncertainty for each data set.  For range measurements, the uncertainty is given in meters of one-way path, while for Doppler data the fractional frequency uncertainty $\Delta\nu/\nu$ is reported.  MGS stands for the Mars Global Surveyor.  Fly-by refers to spacecraft that passed close to Jupiter and Saturn (see text for details).  For reference, the orbital semi-major axes of the planets are given in Table \ref{tab:precSources}.  The line-of-sight velocities of Mercury and Venus, as seen from the Earth, can reach 35 km~s$^{-1}$ and 15 km~s$^{-1}$, respectively.}
\begin{ruledtabular}
\begin{tabular}{lllrrr}
Planet  & Type           & Target     & Date Range        & $N_{obs}$ & $\sigma_{obs}$ (m or $\frac{\Delta\nu}{\nu}$) \\
\hline
Mercury & Radar range    & Surface    & 07/1969 -- 08/1997 &   7,880   &   200     \\
        & Radar Doppler  & Surface    & 07/1969 -- 03/1974 &     174   &   $(5-20)\times10^{-11}$     \\
Venus   & Radar range    & Surface    & 07/1969 -- 07/1982 &   5,466   &   200     \\
        & Radar Doppler  & Surface    & 07/1969 -- 05/1977 &     340   &   $(0.2-20)\times10^{-11}$     \\
% Mariner 9 was the first spacecraft to orbit another planet, 
% narrowly beating Soviet Mars 2 and 3 which both arrived within 
% a month.  Still in Martian orbit.  Should remain there, in a
% stable orbit until 2022
Mars    & Radio tracking & Mariner 9  & 11/1971 -- 10/1972 &     185   &   500     \\ % m9
% Viking 1 and 2 were missions each with an orbiter and a lander
% from 7/1976-8/1980 2 freq tracking of the orbiters was used 
% to correct lander ranges for the plasma propagation.  
% After 8/1980 the 2 freq tracking was not available and so the 
% range uncertainties grew accordingly
        & Radio tracking & Viking 1 \& 2 & 07/1976 -- 11/1982 &   3,872   &    15-40 \\ % rng and postomc lander
%        & Radio tracking & Viking 1+2 Landers & 07/1976 -- 08/1980 &   3,107   &    15   \\ % rng and postomc lander
%        & Radio tracking & Viking 1 Lander  & 08/1980 -- 11/1982 &     765   &    40     \\ % postomc lander
        & Radio tracking & Pathfinder & 07/1997 -- 09/1997 &      90   &    40     \\
        & Radio tracking & MGS        & 02/1999 -- 06/2005 & 124,856   &    12     \\
        & Radio tracking & Odyssey    & 02/2002 -- 06/2005 & 142,416   &     6     \\
% Jupiter Fly-bys
%   Pioneer 10 12/ 4/73
%   Pioneer 11 12/ 3/74
%   Voyager  1  3/ 5/79
%   Voyager  2  7/10/79
Jupiter & Radio tracking & Fly-by &  1973, 1974, 1979 &       4   & 4,500    \\
% Saturn Fly-bys
%   Voyager 1  11/13/80
%   Voyager 2   8/26/81
Saturn  & Radio tracking & Fly-by &  1980, 1981        &       2   & 4,500   
\end{tabular}
\end{ruledtabular}
\end{table*}
%%%%%%%%%%%%%%%%%%%%%%%%%%%%%%%%%%%%%%%%%%%%%%%%%%%%%%%%%%%%%%%%%%%%%%%%%%%%%%%%

\section{Results}
Using the input data described above, and allowing for a single common anomalous precession for the planets Mercury, Venus, Earth, Mars, Jupiter and Saturn, the PEP code derived a universal precession rate of $(d\omega/dt)_{obs} = 0.001 \pm 0.002$ arcseconds per century (1$\sigma$ statistical), consistent with no anomalous precession.  In Section~\ref{verification}, however, we show that this statistical uncertainty overestimates the sensitivity to precession by a factor of 11.  Over 35 years (the approximate duration of our data set) a precession rate of 0.002 arcseconds per century would have perturbed the orbit of Mercury and Mars by $\approx$100 m and $\approx$500 m, respectively, leading to observable range perturbations, $\Delta\rho(t)$, of $\approx$2 m and $\approx$1 m.  For comparison, an independent analysis of planetary data by Pitjeva \cite{pitjevaRelativisticEffects2005}, in which each planet was allowed to precess independently, derived $1\sigma$ statistical uncertainties in the estimates of the planetary perihelia of 5.0, 300, 0.4 and 0.5 milliarcseconds per century for Mercury, Venus, Earth and Mars, respectively.  Thus Pitjeva's tightest constraint on planetary precession, allowing for the independent motion of each body, is five times smaller than our statistical constraint on the uniform precession rate.  We raise a flag of caution regarding the use of published solar system precession constraints based on statistical errors only.  In particular, we show in the next section that the statistical uncertainty of 0.002 arcseconds per century overestimates the true sensitivity of PEP and the data set to a DGP-like precession by a factor of 11.

\section{Estimating the Sensitivity to DGP Precession}
\label{verification}
In order to determine the sensitivity of PEP and the data set to a universal, DGP-like perihelion precession, a series of simulation data sets were generated with the characteristics of our real data (see Table \ref{tab:dataset}), but that also had varying amounts of excess perihelion precession, ranging from zero to 0.2 arcseconds per century, the latter being 100 times the statistical limit reported in the previous section.  The resulting best-fit DGP-like universal precession rates, as determined by PEP, were consistently 11 times smaller than the input precession rates.  This exercise showed that $\sim$90\% of the precession signal was absorbed into other (``nuisance'') model parameters due to the non-linear dependence of the model on the anomalous precession rate.  To account for this dilution, we choose to report an uncertainty on $(d\omega/dt)_{obs}$ that is equal to the amount of input precession required to change the central value of that parameter by one statistical standard deviation (in this case 0.002 arcseconds per century, as reported in the previous section).  Our realistic $1\sigma$ constraint on a common DGP-like precession is therefore:
\[
\left|\frac{d\omega}{dt}\right|_{obs} < 0.02 \text{  arcseconds per century.}
\]

\section{Implications for DGP Gravity}
The universal precession rate for the DGP scenario depends only on $r_c$, the scale above which the graviton can escape the four-dimensional brane and explore the full five-dimensional space.  In order for DGP gravity to explain the observed acceleration rate of the Universe, $r_c$ must be $\approx$5~Gpc.  Our 1$\sigma$ upper limit on the common anomalous planetary precessions in the solar system of $\left|d\omega/dt\right| < 0.02$ arcseconds per century requires that $r_c > 0.13$~Gpc.  This is not yet stringent enough to rule out the DGP scenario as being responsible for the accelerating expansion.  

\section{Conclusion}
Constraints on the periapse precession of solar system bodies can be used to compare solar system observations with speculative gravitational theories, especially those that fall outside of the PPN framework.  We have modified PEP to allow for anomalous precessions of the planetary perihelia, and have performed a least-squares solution of this modified model to a suite of archival radar and radio tracking data, with the requirement that all anomalous precession rates be equal.  At the level of 0.02 arcseconds per century, we find no evidence for a universal precession in excess of the GR prediction.  This result, which places a lower limit of 0.13~Gpc on the cross-over scale in DGP gravity, is a factor of $\sim 40$ away from the prediction of DGP gravity.  We therefore cannot rule out DGP gravity as a mechanism for the accelerating Universe.  It is remarkable, however, that a theory of gravity designed to explain phenomena on cosmological scales has the potential to be tested within the solar system.

We have pointed out that mutually inclined orbits are necessary to distinguish a universal DGP-like precession from a global rotation, and that the observable signature of DGP gravity for interplanetary ranges scales like $\sin^2i$.  We also showed that the statistical uncertainty on the precession rate, determined by a weighted least-squares fit to the observations, overestimates the sensitivity to a DGP-like precession by a factor of 11.  One should therefore exercise caution when interpreting statistical uncertainties from the literature as constraints on physical theories.  

This work can be extended to incorporate the nearly 40 years of high-precision lunar laser ranging observations, including the new, millimeter-precision, LLR data from the APOLLO project.  In addition, the recent successful asynchronous laser transponder \citep{degnan2002} experiment to the MESSENGER spacecraft over 24 million km \cite{smithZuberMessengerTransponder} has demonstrated that $\approx$20 cm precision planetary ranges over interplanetary distances are possible.  Currently, a feasibility study is underway for a project to deploy a laser transponder on the surface of Mars capable of producing millimeter-precision range measurements \citep{murphyPerComm2008}.  Such a high-precision planetary range data set, if extended over several years, would greatly improve the solar system constraints on gravitational physics, and could potentially rule out the DGP scenario.

To facilitate the tie between observation and theory, we encourage others to calculate the relative precession rates of solar system bodies in speculative gravitational theories that do not have a PPN expansion.

\section{Acknowledgements}
We would like to thank the following people for their thoughtful contributions to this work:  Nima Arkani-Hamed, Edmund Bertschinger, Kaisey Mandel, Kenneth Nordtvedt, Irwin Shapiro, James Williams, Matias Zaldarriaga and Phillip Zukin.  We are also grateful for access to the archival observational data used in this project.  J.~B.~R.~B. acknowledges financial support from the National Science Foundation.  We also thank the anonymous referees for suggestions which have improved this work.  

% use bibtex instead of writing all refs by hand...
\bibliography{battat}

\end{document}